\documentclass[sigconf]{acmart}

\usepackage[ruled,vlined,linesnumbered]{algorithm2e}
\usepackage{subfigure}
\usepackage{booktabs}
\usepackage{threeparttable}
\usepackage{pifont}

\newcommand{\cmark}{\ding{51}}%
\newcommand{\xmark}{\ding{55}}%

\AtBeginDocument{%
  \providecommand\BibTeX{{%
    \normalfont B\kern-0.5em{\scshape i\kern-0.25em b}\kern-0.8em\TeX}}}

\copyrightyear{2019} 
\acmYear{2019} 
\setcopyright{acmlicensed}
\acmConference[AISec'19]{12th ACM Workshop on Artificial Intelligence and Security}{November 15, 2019}{London, United Kingdom}
\acmBooktitle{12th ACM Workshop on Artificial Intelligence and Security (AISec'19), November 15, 2019, London, United Kingdom}
\acmPrice{15.00}
\acmDOI{10.1145/3338501.3357371}
\acmISBN{978-1-4503-6833-9/19/11}

\acmSubmissionID{aisec33}

\begin{document}
\fancyhead{}

\title{HybridAlpha: An Efficient Approach for Privacy-Preserving Federated Learning}

\author{Runhua Xu}
\authornote{
The work was done while interning at the IBM Research - Almaden.
}
\orcid{0000-0003-4541-9764}
\affiliation{%
  \institution{University of Pittsburgh}
  \city{Pittsburgh}
  \state{Pennsylvania}
  \postcode{15260}
  \country{United States}
}
\email{runhua.xu@pitt.edu}

\author{Nathalie Baracaldo, Yi Zhou, Ali Anwar and Heiko Ludwig}
\affiliation{%
  \institution{IBM Almaden Research Center}
  \city{San Jose}
  \state{California}
  \country{United States}
}
\email{{baracald, hludwig}@us.ibm.com, {yi.zhou,ali.anwar2}@ibm.com}

\renewcommand{\shortauthors}{Xu et al.}

\begin{abstract}
  Federated learning has emerged as a promising approach for collaborative and privacy-preserving learning. Participants in a federated learning process cooperatively train a model by exchanging model parameters instead of the actual training data, which they might want to keep private.
  However, parameter interaction and the resulting model still might disclose information about the training data used.
  To address these privacy concerns, several approaches have been proposed based on differential privacy and secure multiparty computation (SMC), among others. They often result in large communication overhead and slow training time.
  In this paper, we propose \textit{HybridAlpha}, an approach for privacy-preserving federated learning employing an SMC protocol based on functional encryption. This protocol is simple, efficient and resilient to participants dropping out.
  We evaluate our approach regarding the training time and data volume exchanged using a federated learning process to train a CNN on the MNIST data set. Evaluation against existing crypto-based SMC solutions shows that \textit{HybridAlpha} can reduce the training time by 68\% and data transfer volume by 92\% on average while providing the same model performance and privacy guarantees as the existing solutions.
\end{abstract}



\maketitle

\section{Introduction}
Machine learning (ML) has been widely applied in industry and academia to a wide variety of domains \cite{lecun2015deep, jordan2015machine}. 
While traditional ML approaches depend on a centrally managed training data set, privacy considerations drive interest in decentralized learning frameworks in which multiple participants collaborate to train a ML model without sharing their respective training data sets.
Federated learning (FL) \cite{mcmahan2016communication, konevcny2016federated} has been proposed as a decentralized process that can scale to thousands of participants. Since the training data does not leave each participant's domain, FL is suitable for use cases that are sensitive to data sharing. 
This includes health care, financial services and other scenarios of particular privacy sensitivity or subject to regulatory mandates.

In FL, each participant trains a model locally and exchanges only model parameters with others, instead of the active privacy-sensitive training data. 
An entity called \textit{aggregator} merges the model parameters of different participants. Often, an aggregator is a central entity that also redistributes the merged model parameters to all participants but other topologies have been used as well, e.g., co-locating an aggregator with each participant.
However, this approach still poses privacy risks: inference attacks in the learning phase have been proposed by \cite{nasr2019comprehensive}; deriving private information from a trained model has been demonstrated in \cite{shokri2017membership}; and a model inversion attack has been presented in \cite{fredrikson2015model}.

To address such privacy leakage,  differential privacy \cite{dwork2009differential, dwork2010boosting} has been proposed for a learning framework \cite{abadi2016deep, papernot2018scalable}, in which a trusted aggregator controls the privacy exposure to protect the privacy of the model's output. 
Similarly,  \cite{pettai2015combining} proposes to combine differential privacy techniques and secure multiparty computation (SMC) to support privacy-preserving analyses on private data from different data providers, whereas \cite{bonawitz2017practical} combines  secret sharing and authenticated encryption in a failure-robust protocol for secure aggregation of high-dimensional data.

Inspired from the hybrid methodology \cite{pettai2015combining}, a recent paper \cite{truex2018hybrid} also proposed a hybrid solution that provides strong privacy guarantees while still enabling good model performance.
This hybrid approach combines a \textit{noise-reduction} differential privacy approach with protection of SMC protocol, where the underlying security cornerstone is additive homomorphic encryption, i.e., threshold Paillier system \cite{damgaard2001generalisation}.
Even though the hybrid approach has good model performance and privacy guarantees, it comes with long training time and high data transmission cost and cannot deal with participants dropping out during the FL process.
In Table \ref{tab:overview_cmp}, we summarize existing privacy-preserving approaches from 
the perspectives of threat model, privacy guarantees, and offered features.
We believe a privacy-preserving FL framework should strive for strong privacy guarantees, high communication efficiency, and resilience to changes.
As shown by Table \ref{tab:overview_cmp}, approaches that offer privacy guarantees incur a large number of communication rounds, substantially increasing the training time for FL systems.

\begin{table*}[!t]
    \centering
    \caption{Comparison of privacy-preserving approaches in federated  machine learning framework}
    \label{tab:overview_cmp}
    \begin{threeparttable}
    \begin{tabular}{cccccccc}
        \toprule
            & \multicolumn{2}{c}{Threat Model} & \multicolumn{2}{c}{Privacy Guarantee} & SMC &  \multicolumn{2}{c}{Features} \\
            \cmidrule(ll){2-3} \cmidrule(ll){4-5} \cmidrule(l){6-6} \cmidrule(ll){7-8} 
            Proposed Approach & participant & aggregator & computation & output & type \tnote{$\ast$} & communication  \tnote{$\dagger$} & dynamic participants \\
        \midrule
             Shokri and Shmatikov\cite{shokri2015privacy} & honest & honest & \xmark & \cmark & $-$ & 1 round & \cmark\\
            PATE \cite{papernot2018scalable} & honest & honest & \xmark & \cmark & $-$ & 1 round & $-$\\
            PySyft \cite{ryffel2018generic} & honest & HbC\tnote{$\diamond$} & \cmark & \cmark & HE & 2 rounds\tnote{$\ddagger$} & $-$\\
            Bonawitz et al.  \cite{bonawitz2017practical} & dishonest & HbC\tnote{$\diamond$} & \cmark & \cmark & SS+AE & 3 rounds\tnote{$\ddagger$} & dropout\\
            Truex et al. \cite{truex2018hybrid} & dishonest & HbC\tnote{$\diamond$} & \cmark & \cmark & TP & 3 rounds\tnote{$\ddagger$} & \xmark\\
            \textbf{HybridAlpha (our work)} & dishonest & HbC\tnote{$\diamond$} & \cmark & \cmark & FE & 1 round\tnote{$\ddagger$} & dropout + addition\\
        \bottomrule
    \end{tabular}
    \begin{tablenotes}
        \footnotesize
        \item[$\ast$] ``SS+AE''represents secret sharing techniques with key agreement protocol and authenticated encryption scheme; ``HE'' is homomorphic encryption scheme; ``TP'' is Threshold-Paillier system, a partially additive homomorphic encryption scheme; ``FE'' indicates functional encryption scheme; symbol $-$ indicates non-comparative option.
        \item[$\diamond$] HbC is the abbreviation of Honest but Curious.
        \item[$\dagger$] The count is based on one epoch at the training phase between the aggregator and the participant.
        \item[$\ddagger$] The key distribution communication is not covered here.
    \end{tablenotes}
    \end{threeparttable}
    \vspace{-4mm}
\end{table*}

To address the above-mentioned challenges, we propose  \textit{HybridAlpha}, an efficient approach for privacy-preserving FL.
\textit{HybridAlpha} employs \textit{functional encryption} to perform SMC.
Using \textit{functional encryption}, we define a simple and efficient privacy-preserving FL approach, which also supports a participant group that is changing during the course of the learning process.
We summarize our key contributions as follows:
\begin{itemize}
    \item  We propose \textit{HybridAlpha}, an efficient privacy-preserving FL approach that employs a differential privacy mechanism and defines a SMC protocol from a multi-input functional encryption scheme.
    We adapt such scheme and include additional provisions to mitigate the risk that curious aggregators and colluding participants will infer private information. 
    \item We implement and compare - both theoretically and experimentally - a functional encryption scheme to common, traditional cryptography schemes such as additive homomorphic encryption and its variants, which are typically used for SMC. 
    Our benchmark results will guide future adoption of these cryptosystems in the selection of adequate SMCs for FL.
    \item We describe an implementation of the \textit{HybridAlpha} approach and apply  it to a convolutional neural network. 
    The experimental results on the MNIST dataset show that our \textit{HybridAlpha} framework has efficiency improvements both in training time and communication cost, while providing the same model performance and privacy guarantee as other approaches.
    \item At the same time, we demonstrate a solution to the dynamic participant group issue, which indicates that our proposed framework is robust to participants' dropout or addition.
    We also analyze the security and privacy guarantee of the \textit{HybridAlpha} framework under our defined threat model within a trusted TPA, honest-but-curious aggregators, and partially dishonest participants.
\end{itemize}
To the best of our knowledge, this is the first approach for privacy-preserving federated learning that demonstrates how to  make use of functional encryption to prevent certain inference attacks that would be possible by naively applying this cryptosystem. We demonstrate that our approach has better model performance, stronger privacy guarantee, lower training time and more efficient communication compared to existing solutions.

\noindent\textbf{Organization}. 
The rest of the paper is organized as follows. 
In $\S$\ref{sec:bg}, we introduce  background and preliminaries.
We propose our \textit{HyrbidAlpha} framework and its underlying threat model  in $\S$\ref{sec:propose}.
The evaluation, as well as the security and privacy analysis are respectively presented in $\S$\ref{sec:evaluation} and $\S$\ref{sec:evaluation:sp}.
We discuss related works in $\S$\ref{sec:relatedwork} and conclude the paper in $\S$\ref{sec:conclusion}.

\section{Background and Preliminaries}
\label{sec:bg}
In this section, we introduce the background and explain the underlying building blocks of our proposed framework.

\subsection{Privacy Preserving Federated Learning}
The first FL design aimed to protect the data privacy by ensuring each participant would keep its data locally and uniquely transmit model parameters \cite{mcmahan2016communication}.
Although at first glance it may provide some level of privacy, attacks in the literature have demonstrated that it is possible to infer private information \cite{fredrikson2015model, shokri2017membership, nasr2019comprehensive}.
To fully protect the privacy of the training data from inference attacks,
it is necessary to provide the \textit{privacy of  the computation} and the \textit{output}. 


\noindent \textit{\textbf{Privacy of computation}.}
Malicious participants involved in FL training may have an incentive to infer private information of others.
Messages exchanged with the aggregator contain model updates that leak private information. For instance, if a bag of words is used as embedding to train a text-based classifier, inspecting gradients can help an adversary identify what words where used (e.g., non-zero gradients constitute words used).
SMC protocols can be used to protect inference attacks at training time. These protocols ensure that individual results cannot be exposed while still allowing the computation of aggregated data. 

\noindent \textit{\textbf{Privacy of output}.}
Machine learning models can also leak private information about the training data \cite{fredrikson2015model, shokri2017membership, nasr2019comprehensive}.
Here, adversaries can repeatedly query the model to identify if a particular observation was part of the training data.
To prevent against these attacks, differential privacy has been proposed. In this case, noise is added to the model to protect individual records in the training dataset. 

\noindent \textbf{Existing approaches.}
Table \ref{tab:overview_cmp} presents an overview of privacy preserving approaches (a more thorough description is presented in $\S$\ref{sec:relatedwork}).
Although some of them provide privacy guarantees for the computation and output,
they lack relevant features for FL systems.
In particular, approaches that increase the number of communication rounds can hinder the applicability of FL, as they augment the training time and amount of data exchanged.
For large models such as neural networks,
this is a major concern.

Another important feature should be provided by FL frameworks is the support for
\textit{dynamic participation}.
In some scenarios,
participants may leave the training process at any time, we refer to these as \textit{dropouts}. As shown in Table \ref{tab:overview_cmp}, existing approaches cannot gracefully deal with dropouts and require re-doing an overall training round with new keys. 
New participants may also join the training process at any time. Existing approaches do not provide support for this dynamic flow and require full-re-keying. 

Our proposed \textit{HybridAlpha} reduces significantly the training time by limiting the number of messages exchanged to one by round - substantially less than existing approaches that offer privacy of computation.
In what follows, we present in detail some of the basic building blocks that allow us to achieve this result.




\subsection{Differential Privacy and Multiparty Computation}
\label{sec:bg:dp_smc}

Differential privacy (DP) \cite{dwork2009differential, dwork2010boosting} 
is a rigorous mathematical framework where an algorithm may be described as differentially private if and only if the inclusion of a single instance in the training dataset causes only statistically insignificant changes to the algorithm's output.
The formal definition for DP is as follows:
\begin{definition}
    (Differential Privacy \cite{dwork2009differential}).
    A randomized function $\mathcal{K}$ gives $(\epsilon, \delta)$\textit{-differential privacy} if for all data sets $D$ and $D^{'}$ differing on at most one element, and all $S \subseteq \texttt{Range}(\mathcal{K})$,
    $$\Pr[\mathcal{K}(D) \in S] \le \exp{(\epsilon)}\cdot\Pr[\mathcal{K}(D^{'}) \in S] + \delta.$$
    The probability is taken over the coin tosses of $\mathcal{K}$.
\end{definition}
Note that $\epsilon$\textit{-differential privacy} can be treated as a special case of $(\epsilon, \delta)$\textit{-differential privacy} where $\delta=0$.
To achieve DP, multiple mechanisms
designed to inject noise to the algorithm's output have been proposed. These mechanisms add noise 
proportional to the sensitivity of the output,
a measure of the maximum change of the output resulting by the inclusion of a single data point. 
Popular mechanisms include Laplacian and Gaussian mechanisms, where the Gaussian mechanism for a dataset $D$ is defined as $M(D)= f(D) + N(0, S^2_f \sigma^2)$, where  $N(0, S^2_f \sigma^2)$ is the normal distribution with mean 0 and standard deviation $S_f \sigma$.
By applying the Gaussian mechanism to function $f$ with sensitivity $S_f$ 
satisfies $(\epsilon,\delta)$-differential privacy \cite{dwork2014algorithmic}. 

\noindent\textbf{Noise Reduction through SMC}.
SMC allows multiple parties to compute a function over their inputs, without revealing their individual inputs \cite{bogetoft2009secure, cramer2015secure}.
SMC can be achieved using different techniques such as garbled circuit with oblivious transfer, fully or partially homomorphic encryption, and functional encryption. 

Prior work has shown that it is possible to maintain the same DP guarantee achieved by \textit{local differential privacy} \cite{kairouz2014extremal, qin2016heavy}, i.e., each party adds its own noise independently, and uses SMC to hide individual inputs.
Concretely, using the Gaussian mechanism defined above,
local differential privacy requires each participant to independently add $N(0, S^2_f \sigma^2)$. Considering $n$ parties, the total noise adds up to $n$. However, when applying SMC each participant can add a fraction of the noise 
$N(0, \frac{1}{n}S^2\sigma^2)$ and then use a SMC technique to share the value for aggregation.
As shown in \cite{truex2018hybrid}, this ensures the same DP guarantee while reducing the amount of total noise injected by a factor of $n$.


\subsection{Functional Encryption}
\label{sec:bp:fe}
\textit{HybridAlpha} relies on Functional Encryption (FE), a public-key cryptosystem that allows parties to encrypt their data, meanwhile, an external entity can compute a specific function on the ciphertext without learning anything additional from the underlying plaintext data \cite{boneh2011functional}.
For this purpose, a \textit{trusted third party} (TPA) is used to set up a master private key and a master public key
that will be used to derive multiple public keys to one or more parties who intend to encrypt their data.
Given a function $f(\cdot)$, a functional private key $\text{sk}_{f}$ will be generated by the TPA who holds a master secret key. By processing the function related key $\text{sk}_{f}$ a decryptor can compute $f_{\text{sk}_{f}}(x)$ from the encrypted data, $enc(x)$, without revealing the plaintext $x$.

To satisfy the distributed setting, a variant of functional encryption, i.e., Multi-Input Functional Encryption (MIFE) is proposed in \cite{goldwasser2014multi}, where a functional secret key $\text{sk}_{f}$ can correspond to an n-ary function $f(\cdot)$ that takes multiple ciphertexts as input.
In our federated machine learning framework, we adopt with modifications the MIFE scheme for inner-product proposed in \cite[$\S4.4$]{abdalla2018multi}
due to its computational efficiency.
Unlike other FE schemes, this construction provides a construction that does not require performance intensive pairing operations, and hence reduces the overall run-time.

We now present our construction of a MIFE scheme for federated learning, which derived from the construction in \cite[$\S4.4$]{abdalla2018multi} whose security relies on the Decisional Diffie-Hellman (DDH) assumption.
This MIFE scheme relies on five algorithms, \textit{Setup}, \textit{PKDistribute}, \textit{SKGenerate}, \textit{Encrypt}, \textit{Decrypt}, and three roles: the \textit{third party authority (TPA)}, \textit{participants} and \textit{aggregators}.
Algorithms \textit{Setup, PKDistribute, SKGenerate} are run by \textit{TPA}. 
Each \textit{participant} runs \textit{Encrypt} algorithm to encrypt their model parameters and the \textit{aggregator} runs \textit{Decrypt} algorithm to acquires the average sum of encrypted model parameters.
Notice that comparing to the MIFE construction in \cite{abdalla2018multi}, we add an additional algorithm, i.e., \textit{PKDistribute}, to help deliver the public keys \textit{pk} to \textit{participants}.
As a result, each \textit{participant} will encrypt their data using its own \textit{pk} rather than using the master key \textit{msk} as described in \cite{abdalla2018multi}.
This modification is beneficial because different
parties do not need to share a single master key.

Suppose the inner-product functionality is defined as follows:
$$f((\textbf{x}_{1}, \textbf{x}_{2}, ..., \textbf{x}_{n}),\textbf{y}) = \sum^{n}_{i=1}\sum^{\eta_{i}}_{j=1}(x_{ij}y_{\sum^{i-1}_{k=1}\eta_{k}+j}) \ s.t. |\textbf{y}| = \sum^{n}_{i=1}\eta_{i}, $$
where $n$ denotes the total number of input sources, $\eta_{i}$ is the length of each input vector $\textbf{x}_i$, and $dimension(\textbf{y}) = \sum_{i=1}^n dimension(\textbf{x}_i)$ 
The specific construction of MIFE is defined follows:
\begin{itemize}
    \item $\text{Setup}(1^{\lambda}, \mathcal{F}^{\vec{\eta}}_{n})$:
    The algorithm first generates secure parameters as
    $\mathcal{G}:=(\mathbb{G}, p, g)$ $\leftarrow$ $\text{GroupGen}(1^{\lambda})$, and then generates several samples as $a \leftarrow_{R} \mathbb{Z}_{p}$, $\textbf{a}:=(1,a)^{\intercal}$, $\forall i \in \{1,...,n\}: $ 
    $\textbf{W}_{i} \leftarrow_{R}\mathbb{Z}^{\eta_{i}\times 2}_{p}$, $\textbf{u}_{i} \leftarrow_{R} \mathbb{Z}^{\eta_{i}}_{p}$.
    Then, it generates the \textit{master public key} and \textit{master private key} as 
    $$\textbf{mpk} := (\mathcal{G}, [\textbf{a}]\footnote{Note that $[x]=g^{x}$. Here we implicitly adopt the bracket notation from \cite{escala2017algebraic}, which is somewhat standard in the crypto community.}, [\textbf{Wa}]),\;\; \textbf{msk} := (\textbf{W}, (\textbf{u}_{i})_{i \in \{1,...,n\}}).$$
    \item $\text{PKDistribute}(\textbf{mpk}, \textbf{msk}, \text{id}_{i})$:
    It looks up the existing keys via $\text{id}_{i}$ and returns the \textit{public key} as $\textbf{pk}_{i} := (\mathcal{G}, [\textbf{a}], (\textbf{Wa})_{i}, \textbf{u}_{i})$.
    \item $\text{SKGenerate}(\textbf{mpk}, \textbf{msk}, \textbf{y})$:
    The algorithm first partitions $\textbf{y}$ into $(\textbf{y}_1||\textbf{y}_{2}||...||\textbf{y}_{n})$, where $|\textbf{y}_{i}|$ is equal to $\eta_{i}$.
    Then it generates the function derived key as
    $$\textbf{sk}_{f,\textbf{y}} := ((\textbf{d}^{\intercal}_{i}) \leftarrow (\textbf{y}^{\intercal}_{i}\textbf{W}_{i}), z \leftarrow \sum(\textbf{y}^{\intercal}_{i}\textbf{u}_{i})).$$ 
    \item $\text{Encrypt}(\textbf{pk}_{i}, \textbf{x}_{i})$:
    The algorithm first generates a random nonce $r_{i} \leftarrow_{R} \mathbb{Z}_{p}$, and then computes the ciphertext as 
    $$\textbf{ct}_{i} := ([\textbf{t}_{i}] \leftarrow [\textbf{a}r_{i}], [\textbf{c}_{i}] \leftarrow [\textbf{x}_{i} + \textbf{u}_{i} + \textbf{Wa}_{i}r_{i}]).$$
    \item $\text{Decrypt}(\textbf{ct}, \textbf{sk}_{f,\textbf{y}})$: 
    The algorithm first calculates as follows:
    $$C := \frac{\prod_{i \in [n]}([\textbf{y}^{\intercal}_{i}\textbf{c}_{i}]/[\textbf{d}^{\intercal}_{i}\textbf{t}_{i}])}{z},$$ 
    and then recovers the function result as $f((\textbf{x}_{1}, \textbf{x}_{2}, ..., \textbf{x}_{n}),\textbf{y}) = \log_{g}(C)$.
\end{itemize}
Note that in the federated learning setting, the \textit{aggregator} holds the vector $\textbf{y}$, and each $\eta_{i}$ is set as 1, which indicates the input from each \textit{participant} is a single element instead of the vector as described in the MIFE scheme.

\section{\textit{Hybrid-Alpha} Framework}
\label{sec:propose}
In this section, we present the specific construction of our \textit{HybridAlpha} framework for privacy-preserving federated learning. 
Our framework prevents inference attacks from curious aggregators
and limits the inference power of colluding participants, as detailed later in the threat model. 


\begin{figure*}
    \centering
    \includegraphics[scale=0.52]{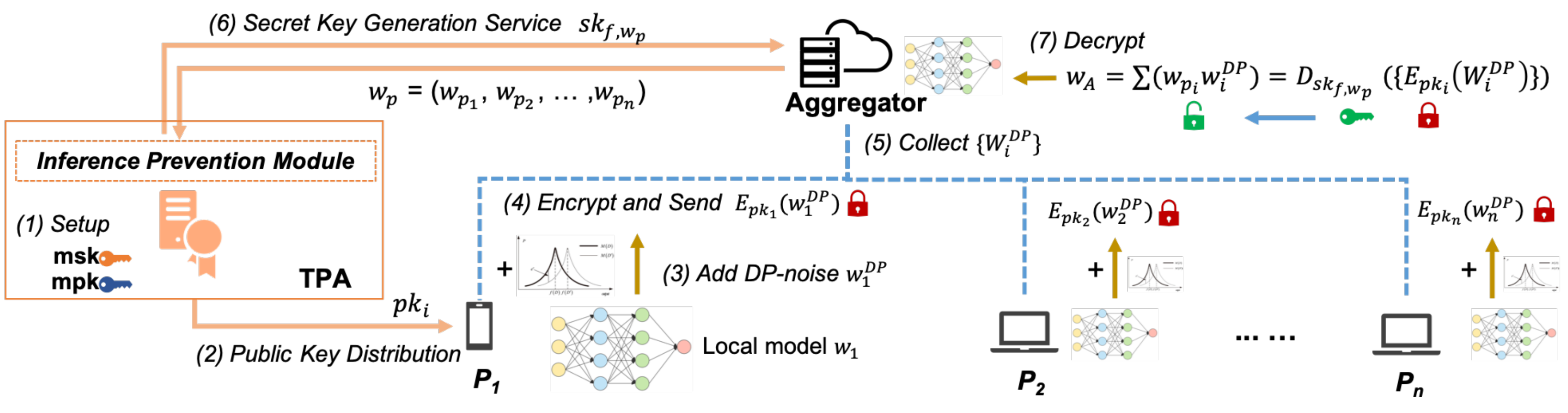}
    \vspace{-4mm}
    \caption{Framework overview of our proposed efficient approach for privacy-preserving federated learning. Note that we only present one epoch here. Each \textit{participant} does the local training based on their owned dataset, and then sends out the model parameters using our proposed efficient privacy-preserving approach.}
    \label{fig:framework}
    \vspace{-4mm}
\end{figure*}

 \figurename\;\ref{fig:framework} presents an overview of \textit{HybridAlpha}.
\textit{Participants} want to collaboratively learn a machine learning model without sharing their local data with any other entity in the system. They agree on sharing only model updates with an aggregator. This entity is in charge of receiving model updates from  multiple participants to build a common machine learning model. 

Participants want to protect their data against any inference attack during the FL process and from the final model.
For this purpose, they join a \textit{HybridAlpha}, 
which has a \textit{Third Party Authority (TPA)}. This entity provides a key management service that initiates the cryptosystem and provides functional encryption keys to all parties.
To prevent potential leakage of information,
\textit{HybridAlpha} also includes an
\textit{Inference Prevention Module}
that limits what type of functional encryption
keys are provided.
This module is designed to ensure that decryption keys cannot be obtained by curious aggregators and to limit potential collusion attacks. We detail this module in $\S$\ref{sec:propose:spec:ip}.

    
    

\subsection{Threat Model}
\label{sec:threat}
We consider the following threat model:

\begin{itemize}
     
    \item \textit{Honest-but-curious aggregator}: 
We assume that the aggregator correctly follows the algorithm and protocols,
    but may try to learn private information inspecting the model updates sent by the participants in the process.
    This is a common assumption \cite{truex2018hybrid, bonawitz2017practical}. 
    
    \item \textit{Curious and colluding participants}: We assume that participants may collude to try to acquire private information from other participants by inspecting the messages exchanged with the aggregator or the final model.  
    \item \textit{Trusted TPA}:
    This entity is an independent agency which is widely trusted by the \textit{participants} and the \textit{aggregator}.
    In real scenarios, different sectors of the economy already have entities that can take such role.
    For instance, in the banking industry, central banks often play a fully trusted role, and in other sectors, a third company such as a service or consultant firm can embody the TPA.
    We also note that assuming such trusted and independent
    agency is a common assumption in existing cryptosystems that have employed the TPA as the underlying infrastructure \cite{boneh2011functional, boneh2001identity, goyal2006attribute}.
    The TPA is in charge of holding the master private and public key. The TPA is also trusted to perform
    public key distribution and function derived secret key generation.
    Similarly, \textit{Inference Prevention Module} is fully trusted.
\end{itemize}

We assume that secure channels are used in all communications, thus, man-in-the-middle and trivial snooping attacks are prevented. 
We also assume a secure key-provisioning procedure such as Diffie-Hellman is in place to protect key confidentiality. 
Finally, attacks that aim to create denial of service attacks or inject malicious model updates are beyond the scope of this paper.


Based on the threat model above, our proposed privacy-preserving framework can ensure that (\romannumeral1) the semi-honest aggregator cannot learn additional information except for the expected output by the differential privacy mechanism, and (\romannumeral2) the malicious colluding participants cannot learn the parameters of other honest participants.  
The specific security and privacy analysis are presented in $\S$\ref{sec:evaluation:sp}.

\subsection{HybridAlpha Detailed Operations }
\label{sec:propose:spec}

We now describe in detail the operations of \textit{HybridAlpha} and
begin by introducing the notation used.
Let $\mathcal{A}$ be the aggregator
and
$\mathcal{S}_{\mathcal{P}}$ be a set of $n$ participants,
where each participant $\mathcal{P}_i$ holds its own dataset $\mathcal{D}_{i}$.
We denote as $\mathcal{L}_{FL}$ the learning algorithm to be trained.

In this section, we first introduce the operations of the framework for non-adversarial settings,
and then explain how additional features are used to protect against the inference attacks defined in the threat model section.

\begin{algorithm}[!t]
\SetAlgoLined
\caption{HybridAlpha }
\label{alg:framework}
\KwIn{
$\mathcal{L}_{FL}$ := Machine learning algorithms to be trained;\\
$\epsilon$ := privacy guarantee; $\mathcal{S}_{\mathcal{P}}$ := set of participants,
where $\mathcal{P}_{i} \in \mathcal{S}_{\mathcal{P}}$ holds its own dataset $\mathcal{D}_i$; 
$N$ := maximum number of expected participants;
$t$ := minimum number of aggregated replies}
\KwOut{Trained global model $\mathcal{M}$}
\SetKwProg{Fn}{function}{}{}
\SetKwRepeat{Do}{do}{while}%
\Fn{TPA-initialization($1^{\lambda}, N, \mathcal{S}_{\mathcal{P}}$)}{
    $\textbf{mpk}, \textbf{msk} \leftarrow \text{MIFE.Setup}(1^{\lambda}, \mathcal{F}^{1}_{N})$ s.t. $N \gg |\mathcal{S}_{\mathcal{P}}|$\;
    \ForEach{$\mathcal{P}_i \in \mathcal{S}_{\mathcal{P}}$ }{
        $\textbf{pk}_i \leftarrow \text{MIFE.PKDistribute}(\textbf{mpk}, \textbf{msk}, \mathcal{P}_{i})$\;
    }
}
\Fn{aggregate($\mathcal{L}_{FL}, \mathcal{S}_{\mathcal{P}}, t$)}{
    \ForEach{$\mathcal{P}_{i} \in \mathcal{S}_{\mathcal{P}}$}{
        asynchronously query $\mathcal{P}_{i}$ with $\text{msg}_{q,i}=(\mathcal{L}_{FL}, |\mathcal{S}_{\mathcal{P}}|)$ \;
    }
    \Do{$|\mathcal{S}_{\text{msg}_{\text{recv}}}| \ge t$ \textbf{and} still in max waiting time}{
        $\mathcal{S}_{\text{msg}_{\text{recv}}} \leftarrow$ collect participant response $\text{msg}_{r,i}$ \;
    }
    \If{$|\mathcal{S}_{\text{msg}_{\text{recv}}}| \ge t$}{
    specify $w_{\mathcal{P}}$ vector; request the $\textbf{sk}_{f,w_{\mathcal{P}}}$ from TPA\;
    $\mathcal{M} \leftarrow \text{MIFE.Decrypt}( \textbf{sk}_{f,w_{\mathcal{P}}}, w_{\mathcal{P}}, \mathcal{S}_{\text{msg}_{\text{recv}}})$\;
    }
    \Return $\mathcal{M}$
}
\Fn{participant-train($\epsilon, t, \text{msg}_{q,i}, \mathcal{D}_{i}, \textbf{pk}_{i}$)}{
    $\mathcal{M}_{i} \leftarrow
    \mathcal{L}_{FL}(\mathcal{D}_{i})$\;
    $\mathcal{M}^{DP}_{i} \leftarrow \text{DP}(\epsilon, \mathcal{M}_{i}, t)$\;
    $\text{msg}_{r,i} \leftarrow \text{MIFE.Encrypt}(\mathcal{M}^{DP}_{i}, \textbf{pk}_{i})$\;
    sends $\text{msg}_{r,i}$ to aggregator\;
}
\end{algorithm}

\subsubsection{Non-adversarial setting}
\textit{HybridAlpha}'s operations under non-adversarial settings are indicated in Algorithm \ref{alg:framework}.
As input, \textit{HybridAlpha} takes the set of participants,
the algorithm used for training, and the differential privacy parameter $\epsilon$.

\textit{HybridAlpha} initiates via the TPA setting up keys in the system.
In particular, the TPA runs the \textit{Setup} and \textit{PKDistribute} algorithms
presented in $\S$\ref{sec:bp:fe},
so that each participant $\mathcal{P}_{i}$ has its own public key $\textbf{pk}_{i}$ (lines 1-5).
We note that \textit{HybridAlpha} allows new participants to join the training process even if it has already started.
To achieve this, the TPA provisions a larger number of keys than the initial set of participants (line 2).
In this way, when new participants join the training process,
they need to acquire the individual public key from the TPA,
and then participate in the learning protocol; 
all this without requiring any changes for other participants.


To begin the learning process,
the aggregator $\mathcal{A}$ asynchronously queries each participant $\mathcal{P}_{i}$ with a query to train the specified learning algorithm $\mathcal{L}_{FL}$ and the number of participant.
Then, the aggregator collects the responses of each party $\mathcal{P}_{i}$ (lines 7-12).

When all responses are received, assuming there is quorum,
$\mathcal{A}$ needs to request a key from the TPA corresponding to the weighted vector $w_{p}$
that will be used to compute the inner product.
That is, the aggregator requests private key $\text{sk}_{f,w_{p}}$ from the TPA based on $w_{p}$.
For computation of average cumulative sum of each participant's model,
$w_{p}$ can be set as $w_{p} = (\frac{1}{n}, \frac{1}{n}, ..., \frac{1}{n}) \;s.t.\;|w_{p}| = n$, where $n$ is the number of received responses.
Then, $\mathcal{A}$ updates the global model $\mathcal{M}$ by applying the decryption algorithm of the MIFE cryptosystem on collected ciphertext set $\mathcal{S}_{\text{msg}_{\text{recv}}}$ and $\text{sk}_{f,w_{p}}$.
Note that here we assume the aggregator $\mathcal{A}$ will get all responses from every participant.
In the case of dropouts, $n$ can be changed so that it reflects the number of participants that are being aggregated.
In the next subsection, we show how \textit{HybridAlpha} provides recommendations to set up $t$ so that the number of allowed dropouts are limited for security reasons.


At the participant side, when a query for training is received by participant $\mathcal{P}_{i}$,
it trains a local model $\mathcal{M}_i$ using its dataset $\mathcal{D}_{i}$.
During the training process\footnote{The differential privacy mechanism depends on the machine learning model being trained. For simplicity, in Algorithm \ref{alg:framework} we show the noise added after the training process takes place.
However, we note that some DP mechanisms add noise during the training process e.g., to train a neural network with the DP mechanism in \cite{abadi2016deep}}, the participant adds differential privacy noise to the model parameters according to the procedure presented in $\S$\ref{sec:bg:dp_smc}.
Finally, $\mathcal{P}_{i}$
encrypts the resulting noisy model using the MIFE encryption algorithm and sends it to the aggregator (lines 18-22).

\subsubsection{Inference Prevention Module}
\label{sec:propose:spec:ip}
In our threat model, we assume 
an \textit{honest-but-curious} aggregator
 that tries to infer private information during the training process.
 We consider multiple potential attacks where the aggregator manipulates the weighted vector to perform inference.

In particular, suppose that $\mathcal{A}$ wants to infer the model of $\mathcal{P}_{i}$.
$\mathcal{A}$ can try to launch an inference attack to obtain the model updates of participant $k$
by setting the weighted vector as follow:
$$
    \forall k \in \{1,...,n\}: \textbf{w}^{'}_{p} = \left(
    \begin{array}{l}
        w_{p_{k}} = 1, \text{if}\; k = i \\
        w_{p_{k}} = 0, \text{if}\; k \neq i
    \end{array}\right)
$$

If a malicious aggregator is allowed a key to perform the inner product of this vector with the model updates,
the model updates of target user $k$ would become visible;
this follows
because $w^{'}_{p}$
zeros-out the model updates of all other participants except for the ones sent by target participant $k$.
If this is not avoided, $\mathcal{A}$ would acquire $w_{i} + \frac{1}{n}N(0,S^2\sigma^2)$ 
as the decryption result of the MIFE cryptosystem.
Here, the reduced noise $\frac{1}{n}N(0,S^2\sigma^2)$ does not provide the expected privacy guarantee to protect $\mathcal{M}_{i}$ of $\mathcal{P}_{i}$ because each honest participant is injecting noise, assuming its model update is aggregated privately with other $n$ participants.

An honest but curious aggregator may also try to create a smaller weighted vector to exclude a subset of participants from the aggregation process.
In the worst case, the malicious aggregator would try to shrink the weighted vector to include one single participant to uniquely ``aggregate'' the model updates of that participant.

Following this same attack vector, a malicious aggregator colluding with dishonest participants may try to build a $w_p$ vector such that:
(\romannumeral1) a target participant model update is included in the vectors;
(\romannumeral2) all other honest participants model updates are not aggregated, and
(\romannumeral3) updates of dishonest participants are included in the aggregation process.
Since the aggregator is colluding with the dishonest participants included in the aggregation process and only the target participant is included in the aggregation, the model update of the target participant is easily reconstructed (its the single unknown variable in the average equation).

To prevent such inference attacks, we propose an additional component called \textit{Inference Prevention Module} collocated with the TPA.
This module intercepts and inspects
requests for private keys for given weighted vectors to
prevent a curious aggregator from obtaining a key that will allow for an inference-enabling inner product.

To this end, the Inference Prevention Module takes as input a parameter
$t$ that defines a threshold on the number of non-colluding participants,
where $t\geq \frac{n}{2}+1$, that is more than half of the participants should not be colluding.
By running Algorithm \ref{alg:ip_filter} and
using parameter $t$, it is possible to prevent the attacks previously described.
In particular, the Inference Module enforces that keys are only provided to weighted vectors that have at least $t$ non-zero elements and that the weight for each included model update is the same.

Threshold $t$ has an impact on the number of dropouts allowed by the system.
Mainly, it helps set up the minimum quorum of participants replying to the system.
\textit{HybridAlpha} allows a limited number of participants to dropout without requiring any re-keying; only the weighted vector sent by the aggregator
needs to be updated by uniquely including the weights of model updates received.

We also note that $t$ has an impact on how much differential privacy noise is added by each participant to achieve a pre-defined $\epsilon$.
Concretely, the number of aggregated replies is always at least $t$,
so as explain in $\S$\ref{sec:bg:dp_smc}, the noise can be adapted to always account for $t$ non-colluding participants contributing to the average, 
e.g., $N(0, \frac{1}{t}S^2\sigma^2)$.
For this purpose, $t$ needs to be communicated among all participants and the aggregator.

\noindent\textbf{Underlying ML models of \textit{HybridAlpha}}. For simplicity we only use the neural networks as the underlying ML model in our FL framework for illustration and evaluation, however, the our \textit{HybridAlpha} supports various ML algorithms.
As functional encryption enables the computation of any inner-product based operation, any model that can be trained through a stochastic gradient descent (SGD)-based algorithm can be trained via our proposed \textit{HybridAlpha}; models in this pool include SVMs, logistic regression, linear regression, Lasso, and neural networks, among others.
Other models such as decision trees and random forests which require aggregating counts from each participant can also be trained by considering the counts sent to the aggregator as a vector.

\begin{algorithm}[!t]
    \SetAlgoLined
    \caption{Inference prevention filter}
    \label{alg:ip_filter}
    \KwIn{$\textbf{w}_{p}$:=A weighted vector to be inspected for inference attacks; 
    $t$:= threshold of minimum number of dropouts and expected number of non-colluding participants}
    \SetKwProg{Fn}{function}{}{}
    \Fn{inference-prevention-filter($\textbf{w}_p$, $t$)}{
        $c_{nz} \leftarrow$ count the non-zero element in $\textbf{w}_p$\;
        \lIf{$c_{nz} < t$}{
            \Return "invalid $w_{p}$"
        }
        \ForEach{non-zero $w_{p_{i}} \in \textbf{w}_p$}{
            \lIf{$w_{p_{i}} \neq \frac{1}{c_{nz}}$}{\Return "invalid $w_{p}$"}
        }
        forward $\textbf{w}_{p}$ to the TPA\;
    }
\end{algorithm}


\section{Evaluation}
\label{sec:evaluation}

In this section we perform a detailed evaluation of our proposed approach to answer the following questions:

\begin{itemize}
    \item How does \textit{HybridAlpha} perform theoretically when compared to existing techniques that have \textit{similar} threat models? More specifically, how many crypto-related operations can be reduced by using \textit{HybridAlpha}?
    \item How does our proposed SMC perform under benchmarking? How does precision setting impact computation time compared with existing techniques? What impact do different numbers of participants have?
    \item  How does \textit{HybridAlpha} compare to existing techniques in terms of performance efficiency?
\end{itemize}

\begin{figure*}
    \centering
    \includegraphics[scale=0.55]{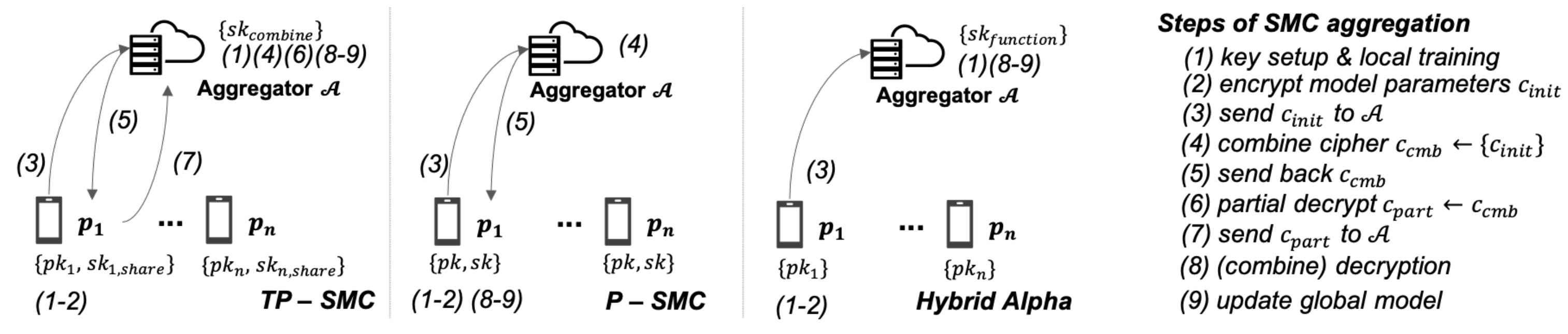}
    \vspace{-4mm}
    \caption{Illustration of aggregation via different crypto-based SMC solutions.}
    \label{fig:smc_overview}
    \vspace{-3mm}
\end{figure*}

\subsection{Baselines and Theoretical Analysis}

We compare the proposed \textit{HybridAlpha} with two state of the art private-federated learning approaches:
\cite{truex2018hybrid} and \cite{ryffel2018generic}, which use different SMC techniques.
A graphical overview and comparison of these baselines can be found in \figurename\;\ref{fig:smc_overview},
the steps performed by each approach are defined in this figure. We will use this notation to report our results.
Additionally, we provide a brief description of our \textbf{baselines}:
\begin{itemize}
    \item We refer to the first baseline as \textbf{TP-SMC} \cite{truex2018hybrid}.
    This FL approach uses a threshold-based homomorphic cryptosystem that allows for a trusted parameter $t$ that specifies the number of participants that are trusted not to collude.
    
    \item We refer as \textbf{P-SMC} to our second baseline which is inspired by PySyft \cite{ryffel2018generic}, an opensource system that uses SPDZ protocol \cite{damgaard2013practical,damgaard2012multiparty}.
    This construct supports homomorphic addition and multiplication. Because the SGD aggregation only requires addition, we opted for a additive homomorphic approach for the comparison, thus, the results reported for this baseline are representative yet faster than PySyft.  
\end{itemize}

We note that the contrasted approaches follows a \textit{similar threat model} to \cite{truex2018hybrid} with a honest-but-curious aggregator, and potentially colluding and malicious participants.
However, they differ in the assumption of a TPA.
We therefore, show how making use of a TPA, \textit{HybridAlpha}
can significantly reduce the training time of machine learning models. \\

\noindent\textbf{Theoretical Comparison}.
We now theoretically compare the crypto-related communication steps associated with the contrasted approaches. 
Suppose that there are $n$ participants and $m$ aggregators in the FL framework, and the threshold for decryption of Threshold-Paillier cryptosystem is $t$.
As shown in Table \ref{tab:efficiency},
In total, \textit{HybridAlpha} reduces $m(n-1)$ and $m(2t-1)$ operations compared to P-SMC and TP-SMC solutions, respectively.
This is achieved because \textit{HybridAlpha} doesn't require sending back encrypted aggregated model updates to the participants for decryption

In the following, we also provide the details of experimental results in $\S$\ref{sec:eval:res}.
The experimental results are consistent with the theoretical analysis.

\begin{table}
  \caption{The number of crypto-related operations required for each solution.}
  \vspace{-2mm}
  \label{tab:efficiency}
  \begin{threeparttable}
  \begin{tabular}{clll}
    \toprule
    Communication & TP-SMC & P-SMC & \textit{HybridAlpha} \tnote{$\ast$} \\
    \midrule
    Step (1) & $n$ & $n$ & $n+m$ \\
    Step (3) & $n\times m$ & $n\times m$ & $n\times m$  \\
    Step (5) & $m\times t$ & $n\times m$ & - \\
    Step (7) & $t \times m$  & - & - \\
    TOTAL   &  $2mt+mn+n$ & $2mn+n$ & $mn+m+n$ \\
    \bottomrule
    \end{tabular}
   \end{threeparttable}
   \vspace{-5mm}
\end{table}

\subsection{Experimental Setup}
To benchmark the performance of \textit{HybridAlpha},
we train a convolutional neural network (CNN) with the same topology as the one used in \cite{truex2018hybrid} to classify the publicly available MNIST dataset of handwritten digits \cite{yann2010mnist}.
The CNN has two internal layers of ReLU units, and a softmax layer of ten classes with cross-entropy loss. The first layer contains 60 neurons and the second layer contains 1000 neurons.
The total number of parameters of this CNN is 118110. 
We also use the same hyperparameters reported in previous work:
a learning rate of 0.1, a batch rate of 0.01. and for differential privacy we use
a norm clipping of 4.0, and an epsilon of 0.5.
We used noise-reduction method as in \cite{truex2018hybrid} as differential private mechanism.
We run experiments for 10 participants,
and each participant was randomly assigned 6,000 data points from the MNIST dataset. For model quality, we used the pre-defined MNIST test set.
Our implementation uses Keras with a Tensorflow backend.

\noindent\textbf{\textit{Cryptosystems Implementation}}
We implement the contrasted cryptosystems in python based on
the opensource integer group of the Charm framework \cite{charm13}.
Charm uses a hybrid design, where the underlying performance-intensive mathematical operations are implemented in native C modules, i.e., the GMP library \footnote{The GNU Multiple Precision Arithmetic Library (https://gmplib.org/).}, while cryptosystems themselves can be written in a readable, high-level language.
Even though there exists Paillier implementation including its threshold variant using other programming languages, we re-implement them in a unified platform 
to allow for fair benchmarking and to enable easy integration with python-based machine learning frameworks such as Keras and Tensorflow. 

In our implementation, we incorporated the following accelerating techniques. In \textit{HybridAlpha}, as presented in $\S$\ref{sec:bp:fe}, the final step of MIFE decryption is to compute the discrete logarithm of an integer, which is a performance intensive computation.
An example would be how to compute $f$ in $h = g^{f}$, where $h, g$ are big integers, while $f$ is a small integer.
To accelerate the decryption, we use a hybrid approach to solve the discrete logarithm problem. 
Specifically, we setup a hash table $T_{h,g,b}$ to store $(h, f)$ with a specified $g$ and a bound $b$, where $ -b \le f \le b$, when the system initializes. 
When computing discrete logarithms, the algorithm first looks up $T_{h,g,b}$ to find $f$, where the complexity is $\mathcal{O}(1)$.
If there is no result in $T_{h,g,b}$, the algorithm employs the traditional \textit{baby-step giant-step} algorithm \cite{shanks1971class} to compute $f$, where the complexity is $\mathcal{O}(n^{\frac{1}{2}})$ .

The second acceleration method we implemented modifies the encryption and decryption algorithms to allow for a one-shot encryption call of a tensor. Here, each generated random nonce is applied to the whole tensor instead of a single element.
We note that a further performance enhancement technique that could be used is parallelizing the encryption/decryption implementation, however, we did not include this enhancement.


\noindent\textbf{\textit{Experimental Setup}}.
All the experiments are performed on 
a  2 socket, 44 core (2 hyperthreads/core) Intel Xeon E5-2699 v4 platform with 384 GB of RAM. Note that the FL framework is simulated
(not run on the real distributed environment), hence the network latency issues are not considered in our experiment.
However, we report a comparison of data transfer by contrasted approaches.

\subsection{Experimental Results}
\label{sec:eval:res}
Here, we first present the benchmark result of three contrasted approaches, and then show the experimental efficiency improvement.

\noindent\textit{\textbf{Impact of Floating Point Precision}}.
The parameters of a neural network (weights) are represented as floating point numbers.
However, cryptosystems take them as input integers.
Hence, the floating point parameters should be represented and encoded into integers.
The \textit{precision number} denotes the number of bits used after the decimal point of a floating point number.
In Table \ref{tab:precision} we present the impact of the precision on the computation time of each crypto-based SMC.
Based on our experimental results, the precision setting has no significant impact on operation time of each cryptosystem.
To be specific, the time cost of encryption, decryption, and other ciphertext computations in each cryptosystem is stable, respectively, of length of the integer.

For encryption, the average time cost of 10 participants on 118110 gradients for \textit{HybridAlpha} is around 4 seconds, while the time cost of P-SMC and TP-SMC under the same setting is about 35 seconds.
For decryption, under the same setting, the cost time of HybridAlpha is about 30 seconds, while the time cost of P-SMC and TP-SMC are 31 and 88 seconds, respectively.
Note that the decryption time of TP-SMC includes the share decryption by part of participants and the final combination decryption by the aggregator, without considering network latency of transmitting the partial decrypted ciphertext.
We can conclude that our proposed approach has significant advantages on both encryption/decryption time cost comparing to P-SMC and TP-SMC solutions. 

Finally, the number of decimal points used in the conversion impacts the overall accuracy of the trained model. In the remaining of the experiments, we used 6-digits which allows for good model and training time performance.

\begin{table*}
  \caption{The impact of precision on computation time (s) of three SMC approaches.}
  \vspace{-2mm}
  \label{tab:precision}
  \begin{threeparttable}
  \begin{tabular}{cccccccccc}
    \toprule
    & \multicolumn{4}{c}{TP-SMC \tnote{$\triangleright$}} & \multicolumn{3}{c}{P-SMC}  & \multicolumn{2}{c}{\textit{HybridAlpha}} \\
    \cmidrule(ll){2-5} \cmidrule(lll){6-8} \cmidrule(llll){9-10}
    precision & $\text{enc}_{\text{avg}}$ & $\text{ct}_{\text{fuse}}$ & $\text{dec}_{\text{share,avg}}$ & $\text{dec}_{\text{combine}}$ & $\text{enc}_{\text{avg}}$ & $\text{ct}_{\text{fuse}}$ & $\text{dec}$ &  $\text{enc}_{\text{avg}}$ & $\text{dec}$ \\
    \midrule
    2 & 35.120 & 2.586 & 61.080 & 28.465 & 35.752 & 2.269 & 32.042 & 4.157 & 30.075 \\
    3 & 35.675 & 2.604 & 61.929 & 28.202 & 35.725 & 2.369 & 31.574 & 4.158 & 30.512 \\
    4 & 35.841 & 2.571 & 60.832 & 28.324 & 35.821 & 2.387 & 31.856 & 4.110 & 29.865 \\
    5 & 35.767 & 2.635 & 60.369 & 28.816 & 35.857 & 2.493 & 31.625 & 4.075 & 30.149 \\
    6 & 35.724 & 2.578 & 60.326 & 28.286 & 35.985 & 2.532 & 31.587 & 4.095 & 30.803\\
    \bottomrule
    \end{tabular}
    \begin{tablenotes}
        \item[$\triangleright$] The threshold parameter of Threshold-Paillier encryption system is set to half the number of participants.
    \end{tablenotes}
    \end{threeparttable}
    \vspace{-3mm}
\end{table*}

\noindent\textit{\textbf{Impact of Number of Participants}}.
We also measure the impact of the number of participants on the time cost for each crypto operation.
The experimental results are shown in Table \ref{tab:participant}.
We see two different trends on the participant and on the aggregator side. 
At the participant side, the encryption and decryption runtime stays the same for all of the evaluated approaches as the number of participants increases.
In contrast, on the aggregator side,
the time cost of ciphertext multiplication increases almost linearly with the increase in the number of participants
(shown in italicized numbers in Table \ref{tab:participant}). 
However, we note a significant difference between \textit{HybridAlpha} and TP-SMC.
For \textit{HybridAlpha} the decryption time increases approximately linearly with the increase of participants, while for 
TP-SMC, the decryption time increases exponentially as the number of participants increases.

\begin{table*}
  \caption{The impact of participant \# on computation time (s) of three SMC approaches.}
  \vspace{-2mm}
  \label{tab:participant}
  \begin{threeparttable}
  \begin{tabular}{cccccccccc}
    \toprule
    & \multicolumn{4}{c}{TP-SMC \tnote{$\triangleright$}} & \multicolumn{3}{c}{P-SMC}  & \multicolumn{2}{c}{\textit{HybridAlpha}} \\
    \cmidrule(ll){2-5} \cmidrule(lll){6-8} \cmidrule(llll){9-10}
    participants & $\text{enc}_{\text{avg}}$ & $\text{ct}_{\text{fuse}}$ & $\text{dec}_{\text{share,avg}}$ & $\text{dec}_{\text{combine}}$ & $\text{enc}_{\text{avg}}$ & $\text{ct}_{\text{fuse}}$ & $\text{dec}$ &  $\text{enc}_{\text{avg}}$ & $\text{dec}$ \\
    \midrule
    6 & 35.968 & \textit{1.375} & 60.555 & \textit{22.184} & 35.934 & \textit{1.332} & 31.616 & 4.241 & \textit{20.246}\\
    8 & 35.375 & \textit{1.843} & 60.820 & \textit{23.980} & 36.039 & \textit{1.859} & 31.611 & 4.092 & \textit{25.349}\\
    10 & 35.693 & \textit{2.358} & 60.988 & \textit{28.401} & 36.847 & \textit{2.611} & 32.197 & 4.077 & \textit{31.782}\\
    12 & 35.685 & \textit{2.759} & 60.947 & \textit{34.684} & 36.142 & \textit{2.959} & 31.588 & 4.091 & \textit{36.884}\\
    14 & 35.688 & \textit{3.215} & 60.965 & \textit{39.838} & 35.932 & \textit{3.330} & 31.503 & 4.126 & \textit{42.683}\\
    16 & 35.721 & \textit{3.694} & 60.917 & \textit{46.849} & 36.533 & \textit{4.481} & 32.020 & 4.059 & \textit{47.435}\\
    18 & 35.683 & \textit{4.170} & 60.879 & \textit{53.441} & 36.628 & \textit{5.368} & 32.996 & 4.594 & \textit{56.519}\\
    20 & 35.697 & \textit{4.764} & 60.816 & \textit{97.224} & 36.743 & \textit{5.765} & 31.923 & 4.147 & \textit{59.823}\\
    \bottomrule
    \end{tabular}
    \begin{tablenotes}
        \item[$\triangleright$] The threshold parameter of TP-SMC is set to half the number of participants.
    \end{tablenotes}
    \end{threeparttable}
    \vspace{-3mm}
\end{table*}

Focusing on the TP-SMC, we also evaluate the impact of threshold $t$, 
which indicates the minimum number of participants who are required to do partial decryption.
As shown in Table \ref{tab:threshold}, only the final decryption has significant relationship with threshold $t$.
For the same number of participants, the cost time of decryption increases linearly as the threshold number increase.

\begin{table}
  \caption{Impact of threshold $t$ for TP-SMC on computation time (s).}
  \vspace{-2mm}
  \label{tab:threshold}
  \begin{threeparttable}
  \begin{tabular}{ccccc}
    \toprule
    threshold\tnote{$\dagger$} & $\text{enc}_{\text{avg}}$ & $\text{ct}_{\text{fuse}}$ & $\text{dec}_{\text{share,avg}}$ & $\text{dec}_{\text{combine}}$ \\
    \midrule
    2 & 35.577 & 2.602 & 60.736 & 12.700 \\
    4 & 35.697 & 2.592 & 60.420 & 23.293 \\
    6 & 35.713 & 2.625 & 60.238 & 34.427 \\
    8 & 36.054 & 2.623 & 60.767 & 46.462 \\ 
    10 & 35.880 & 2.626 & 60.650 & 58.293 \\
    \bottomrule
    \end{tabular}
    \begin{tablenotes}
        \item[$\dagger$] The total participants number is set to 10 and the precision number is set to 6.
    \end{tablenotes}
    \end{threeparttable}
    \vspace{-3mm}
\end{table}

\noindent{\textbf{\textit{Model Quality, Training Time and Data Transmission }}}.
In this experiment, we evaluate the performance of \textit{HybridAlpha} with respect to multiple techniques to perform FL.
In particular, we assess the quality of models produced and the total training time.
The contrasted approaches for this experiment include the following additional baselines:
(\romannumeral1) ``FL-no-privacy'', where the neural network is trained without privacy considerations. This method provides a baseline for maximum possible performance in terms of model quality;
(\romannumeral2) ``Local DP'', where each participant applies differential privacy locally according to \cite{kairouz2014extremal};
(\romannumeral3) for ``TP-SMC'', ``P-SMC'' and ``\textit{HybridAlpha}'', we report the results for two cases: adding differential privacy to protect privacy of the output and without adding differential privacy.
When no differential privacy is added, we use ``TP-SMC no DP'', ``P-SMC no DP'' and ``\textit{HybridAlpha} no DP''.
For privacy-preserving approaches we use an $\epsilon=0.5$.
Finally, our experiments used $t=5$ for \textit{HybridAlpha} and TP-SMC. 
This experiment was run with 10 participants.



\begin{figure*}[!t] 
	\centering 
	\subfigure[F1 score of different approaches as epoch increases]{
		\includegraphics[scale=0.12]{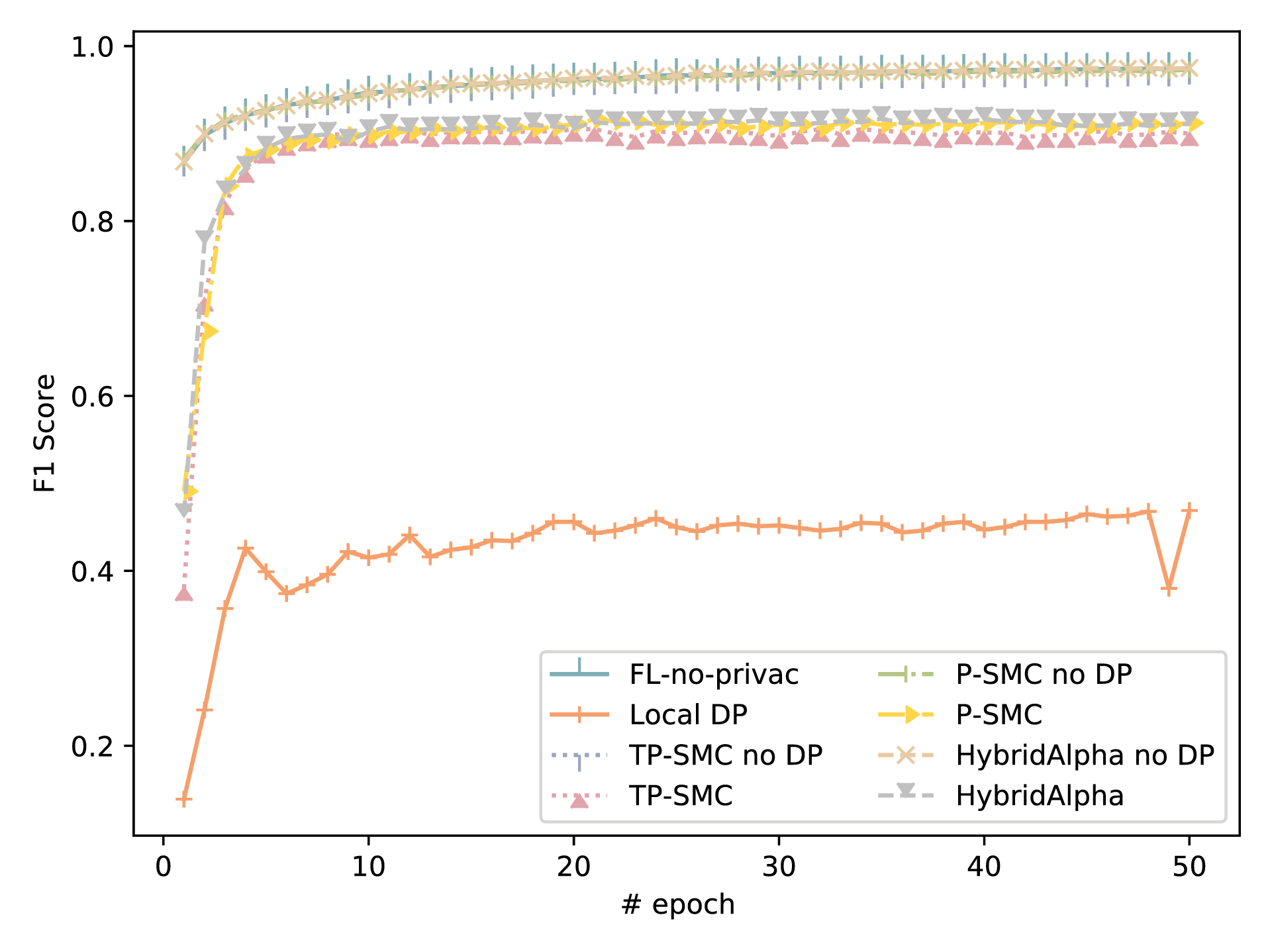}
		\label{fig:efficiency:f1}
		\vspace{-2mm}
	}
	\subfigure[training time of different approaches as epoch increases]{
		\includegraphics[scale=0.12]{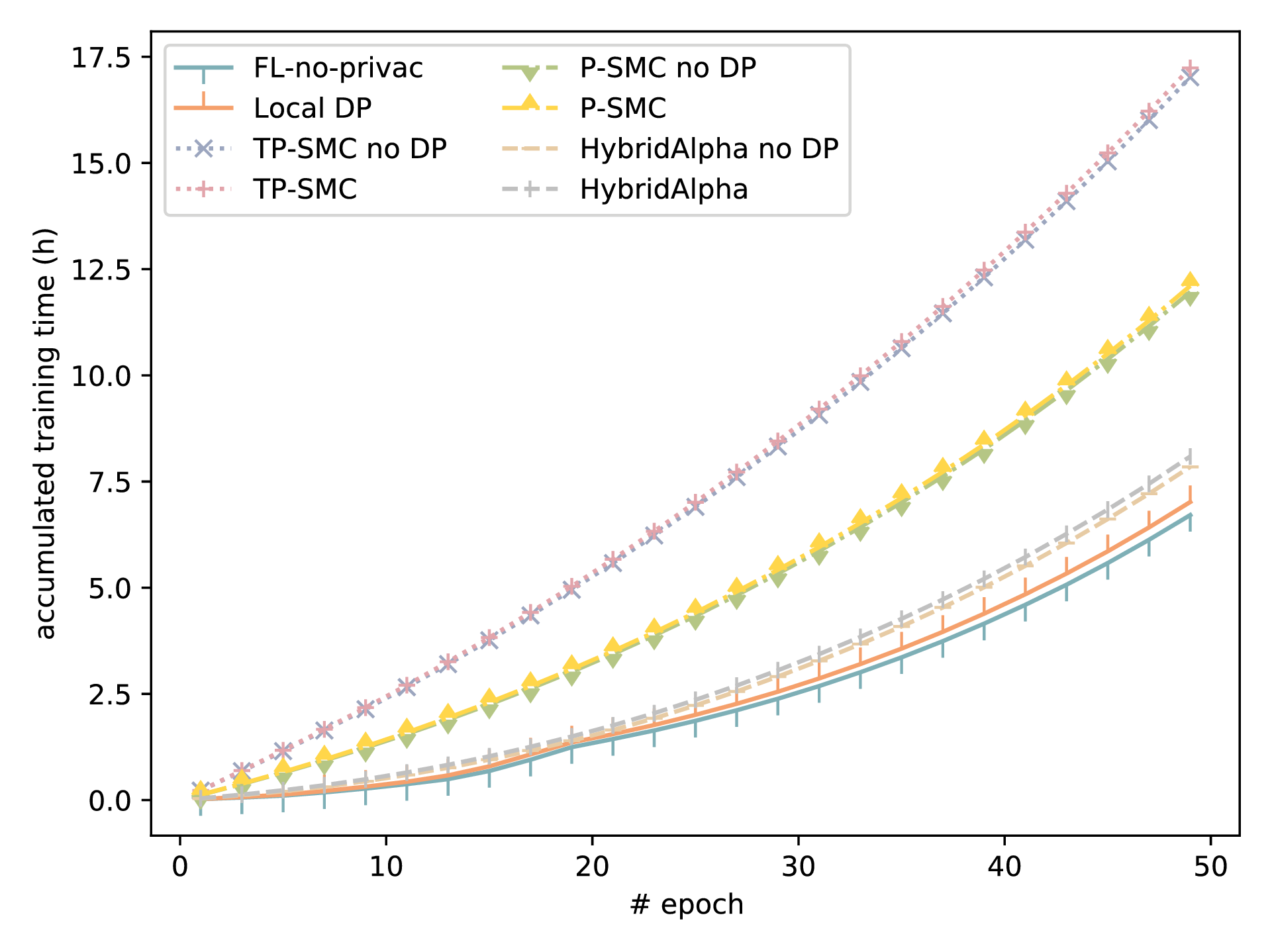}
		\label{fig:efficiency:time}
		\vspace{-2mm}
	}
	\vspace{-5mm}
	\caption{Model quality and time efficiency comparison for multiple FL approaches.}
	\label{fig:efficiency} 
	\vspace{-3mm}
\end{figure*}

To measure quality of model performance,
we report the F1-score (a measure that combines precision and recall) of the resulting models.
The results are presented in \figurename\;\ref{fig:efficiency:f1}.

We see different trends depending on whether a particular approach protects privacy of the computation and of the output.
As expected, approaches that do not protect the privacy of the final model - those that don't inject differential privacy noise- result in a in higher F1-score.
In contrast, ``Local DP'' provides the lowest F1-score due to the high amount of noise injected by each participant.
For approaches that use SMC to uniquely protect the privacy of the computation,
``TP-SMC no DP'', ``P-SMC no DP'' and ``\textit{HybridAlpha} no DP'', we see higher F1-scores than for those that protect the privacy of the output. This shows the price of protecting against the risk of inference on the model.
Finally, we see that approaches that combine differential privacy with SMC are capable of achieving higher F1-scores while protecting the privacy of the input and output.

We now analyze these approaches from the perspective of total training time presented in \figurename\;\ref{fig:efficiency:time}.
As it can be seen, our proposed \textit{HybridAlpha} has very similar training time to ``FL-no-privacy''.
In other words, the training time added by ensuring privacy of the input and output is negligent.
In contrast, we see that the slowest approach is TP-SMC even though we set up $t$ to a conservative 50\% of the entire number of participants in the system.
This result is due to the fact that TP-SMC requires more rounds of communication per global step. The high training time makes TP-SMC suitable for models that require limited number of interactions with the aggregator during training.

\begin{figure}[!t]
    \centering
    \includegraphics[scale=0.09]{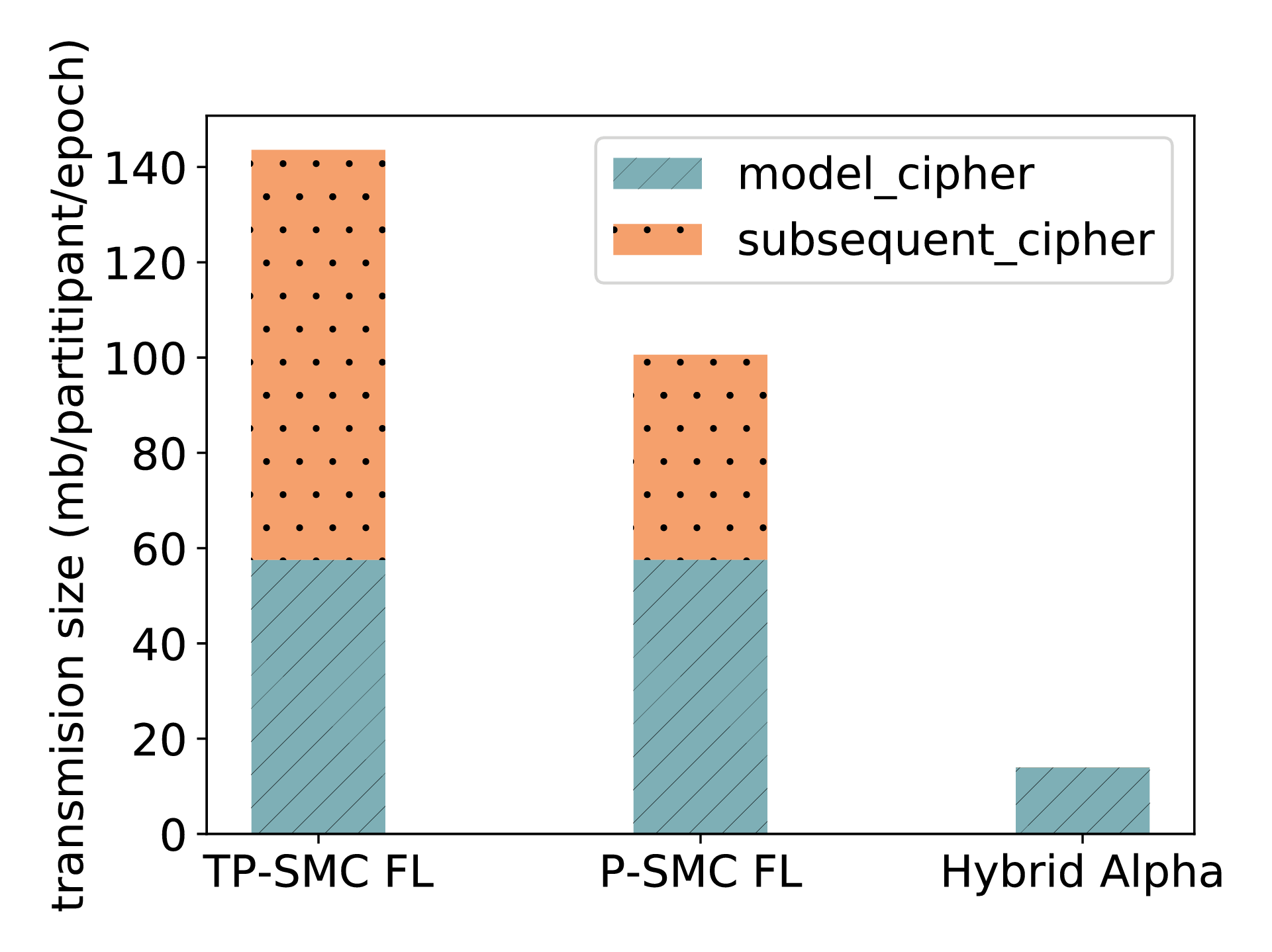}
    \vspace{-5mm}
    \caption{Total transmitted ciphertext size of different approaches for one epoch.}
    \label{fig:cipert_size}
    \vspace{-5mm}
\end{figure}

Beside the efficiency in training time, we also evaluate the efficiency of network transmission by measuring the volume of encrypted parameters transmitted over the network.
In \figurename\;\ref{fig:cipert_size}, we present the total transmitted cipertext size under different crypto-based SMC approaches for one epoch.
The green bar represents initial ciphertext size of model parameters, while the spotted orange bar indicates the size of subsequent ciphertext, including multiplied cipher, and partially decrypted ciphers.
We can see that \textit{HybridAlpha} provides the lowest transmission rate because it only performs one round of communication on encrypted data without any subsequent ciphertext transmission.
Also, our proposed approach has smaller ciphertext size of initial parameters compared to contrasted approaches. 

\section{Security and Privacy Analysis}
\label{sec:evaluation:sp}
We analyze the security and privacy of our proposed framework from three different perspectives: 
security offered by MIFE scheme, privacy guarantees of the framework, and prevention for different types of inference attacks.

\subsection{Security of the Cryptographic Approach}
\label{sec:evaluation:sp:fe}
The security of MIFE is critical to \textit{HybridAlpha}, since it is the underlying infrastructure of SMC protocol that supports secure aggregation in \textit{HybridAlpha}.
In our adoption of MIFE, we add a \textit{public key distribution} algorithm run by the TPA as a beneficial supplement of the original MIFE scheme proposed in \cite{abdalla2018multi} to make it applicable to our FL framework. 

Specifically, the additional algorithm is only responsible for distributing each participant's respective unique public key $\textbf{pk}_{i}$. 
Unlike the original design of encryption algorithms where each participant encrypts the data using the master secret key $\textbf{msk}$, 
our encryption algorithm uses $\textbf{pk}_{i}$ that is derived from the master keys $\textbf{mpk}$ and $\textbf{msk}$.
However, the core method in the encryption algorithm remains intact, and our design has no impact on other algorithms, e.g., \textit{SKGenerate, Decrypt}.
As a consequence, our adoption of MIFE does not change the security construction in the original MIFE scheme in \cite{abdalla2018multi}. It is then as secure as proved in \cite{abdalla2018multi}.
To avoid redundancy, we do not present the correctness and security proofs to MIFE here, and readers can refer to \cite{abdalla2018multi} for more details.

\subsection{Privacy of FL Framework}
Our proposed FL framework can ensure the privacy of the output model and the aggregation computation.

\subsubsection{Privacy of the Output Model}
We provide $\epsilon$\textit{-differential privacy} guarantee via existing methods presented in previous works, e.g., \cite{truex2018hybrid, abadi2016deep,pettai2015combining}. These papers have shown via theoretical analysis and experimental results that such a mechanism can achieve target privacy along with acceptable performance for the final trained model. 
As a consequence, our proposed framework can also achieve the same privacy guarantee for the output model as demonstrated in \cite{truex2018hybrid,pettai2015combining}.

\subsubsection{Privacy of Computation}
We exploit multi-input functional encryption as the underlying infrastructure for SMC protocol to compute the average of the weights of the participants' local trained models.
As stated in $\S$\ref{sec:evaluation:sp:fe}, the MIFE scheme is secure so that any plaintext under its protection cannot be compromised by malicious attackers.
The MIFE scheme also guarantees that the decryptor, the aggregator in our FL framework, can only acquire the function results, i.e., the average weight, but not the original data, i.e., weights of the participants' local models.

\subsubsection{Inference Attack Prevention}
Next, we consider inference attacks for two adversaries:
(\romannumeral1) a curious aggregator, and (\romannumeral2) malicious or colluding participants.

In $\S$\ref{sec:propose:spec:ip}, we have shown that a curious aggregator can launch an inference attack targeting a specific participant by manipulating the \textit{weighted vector} $w_{p}$ and subsequently requesting the function private key.
To prevent such inference attacks, we add an additional module in TPA to filter
requests for weighted vectors that are maliciously defined to isolate the reply of a single participant. Algorithm \ref{alg:ip_filter} verifies that at least $t$ replies are used for aggregation, because there are at least $t>(n/2)+1$ non-colluding parties;
even if the aggregator colludes with dishonest participants he cannot isolate the reply of a target participant.

Even if an adversary manages to  collect other participants' encrypted data for a possible brute-force attack, this attack is not successful.
In particular, suppose that there exits a malicious participant $\mathcal{P}^{'}_{i}$ with its own public key $\textbf{pk}_{\mathcal{P}^{'}_{i}}$, collected encrypted data $c_{j} = \text{Enc}_{\textbf{pk}_{\mathcal{P}_{j, j\neq i}}}(m_{j})$ from $\mathcal{P}_{j}$, and its own original data set $\mathcal{S}^{'}$.
Here $m_{j}$ is the corresponding plaintext of $c_{j}$, 
and $m_{j}$ and any $ m^{'} \in \mathcal{S}^{'}$ belong to a same integer group. 
The semantic security of the underlying MIFE scheme in our SMC protocol ensures that the adversary $\mathcal{P}^{'}_{j}$ does not have a non-negligible advantage to infer the original data $m_{j}$ compared to the random guess. 
Furthermore, as we assume the existence of at least $t$ honest participants where each participant does not share the same public key for encryption, the colluding participants cannot infer/identify private information using the output of the aggregator with their local models. 

Note that based on the threat model defined in $\S$\ref{sec:threat}, we do not consider the DDoS attack on the aggregator where a malicious aggregator or a outside attacker will interrupt the network or replace a valid update from an honest participant.

\section{Related Work}
\label{sec:relatedwork}
Federated learning was proposed in \cite{mcmahan2016communication, konevcny2016federated} to jointly learn a global model without sharing their data.
Although this provides a basic privacy guarantee because privacy-sensitive data is not transmitted,
it still suffers from inference attacks that use the final model or the model updates exchanged during the FL training to infer private information; examples of such attacks include  \cite{nasr2019comprehensive, fredrikson2015model,shokri2017membership, nasr2018machine}.

To thwart inference attacks, previous work has proposed to add differential privacy during the learning process e.g., \cite{abadi2016deep}.
However, directly applying such approaches in the FL setting results in poor model performance because each party trains its own model.
For this reason, new approaches tailored to FL have been proposed.

Table \ref{tab:overview_cmp} shows the threat model of existing approaches and the privacy guarantees they provide.
Solutions proposed in \cite{papernot2018scalable,shokri2015privacy} rely on trusted aggregators and honest participants.
PATE \cite{papernot2018scalable} proposes a ``teacher-student'' architecture
where teachers train their models over their local datasets.
Subsequently, a fully trusted aggregator is used to build a collective model.
All these approaches differ from \textit{HybridAlpha} in that they rely on a trusted aggregator. 

A number of approaches propose to use SMC \cite{ryffel2018generic, bonawitz2017practical, truex2018hybrid}. 
Among them, different SMC protocols are used to aggregate the global model (see Table \ref{tab:overview_cmp}).
PySyft \cite{ryffel2018generic} employs the SPDZ protocol \cite{damgaard2013practical,damgaard2012multiparty}. Truex et al. \cite{truex2018hybrid} uses a threshold-based partially additive homomorphic encryption \cite{damgaard2001generalisation}.
Bonawitz et al. \cite{bonawitz2017practical} makes use of secret sharing that enables authenticated encryption.
These approaches \cite{ryffel2018generic, truex2018hybrid, bonawitz2017practical} also prevent or can be extended to prevent inference attacks on the final model by adding differential privacy.
The main advantage \textit{HybridAlpha} has over these approaches is the communication efficiency (see Table \ref{tab:overview_cmp}). 
While existing approaches need more than one round of communication between the aggregator and participants (excluding the key setup phase), \textit{HybridAlpha} only incurs a single round.
Hence \textit{HybridAlpha} can be used to train machine learning models faster as demonstrated in the experimental section.

Most of the existing federated learning frameworks only work on the scenario of horizontally partitioned data.
To tackle the issues and challenges in the case of vertically partitioned data, several methods are proposed in \cite{hardy2017private, djatmiko2017privacy, cheng2019secureboost}, which focus on entity resolution and simple machine learning models, like logistic regression.
Such vertically partitioned data cases beyond the research scope in this paper, and will be deferred to future work. 

\section{Conclusion}
\label{sec:conclusion}

Federated learning promises to address privacy concerns and regulatory compliance such as for GDPR \cite{Voigt2017GDPR} and HIPAA \cite{act1996health}.
However, extant approaches addressing privacy concerns in federated learning provide strong privacy guarantees and good model performance at the cost of higher training time and high network transmission.
We propose \textit{HybridAlpha}, a novel federated learning framework to address these issues.
Our theoretical and experimental analysis shows that, compared to existing techniques, on average, \textit{HybridAlpha} can reduce the training time by 68\% and data transfer volume by 92\% without sacrificing privacy guarantees or model performance.
Using \textit{HybridAlpha}, federated learning with strong privacy guarantees can be applied  to use cases sensitive to training time and communication overhead, in particular for models with a large number of parameters.




\bibliographystyle{ACM-Reference-Format}
\bibliography{references}

\end{document}